\documentclass[cameraready]{Interspeech}

\title{Acoustic Prompting via Stage-wise Modulation \\ for Few-Shot Learning in Audio Language Models}

\author{Hyebin}{Cho}
\author{Jaehyuk}{Jang}
\author[correspondingauthor]{Changick}{Kim}
\author[correspondingauthor]{Joon Son}{Chung}

\address{
Korea Advanced Institute of Science and Technology, South Korea
}

\email{\{hyebin.cho, jhyuk, changick, joonson\}@kaist.ac.kr}


\usepackage{comment}
\usepackage{booktabs}
\usepackage{makecell}
\usepackage{tabularx}
\usepackage{array}
\usepackage{amsmath}
\usepackage{amssymb}
\usepackage{graphicx}
\DeclareMathOperator*{\argmax}{argmax}

\usepackage{color, colortbl}
\usepackage[table]{xcolor}
\definecolor{Green}{rgb}{0.2, 0.7, 0.1}
\definecolor{lightgray}{gray}{0.93}
\usepackage{kotex}

\begin{document}

\maketitle 
\begin{abstract}
    Audio-Language Models (ALMs) have shown remarkable success in zero-shot audio classification by aligning audio waveforms with text. Recent efforts to improve downstream performance focus on learning optimal text prompts. However, previous approaches focus on the text encoder, leaving the potential of learnable prompts within the audio encoder unexplored. In this paper, we propose a novel framework that introduces trainable prompts into the audio encoder to capture task-specific acoustic features. We demonstrate that integrating audio-side prompt learning with existing text-side approaches enhances few-shot adaptation. Through extensive experiments across 11 datasets show that integrating our method as a plug-and-play module alongside existing text prompt tuning generally leads to performance improvements. These findings suggest that explicitly modulating the audio representation space effectively complements text-only prompting approaches. The code is available at \url{https://github.com/hyebin-c/aspl}.
\end{abstract}

\section{Introduction}
Recent advances in large-scale multimodal foundation models have fundamentally transformed the landscape of audio understanding. In particular, Audio-Language Models (ALMs) have emerged as powerful tools, enabling highly effective open-vocabulary recognition across diverse acoustic domains. By aligning audio waveforms with natural language descriptions in a shared embedding space, these models, such as Contrastive Language–Audio Pretraining (CLAP) \cite{clap}, AudioCLIP \cite{audioclip}, and Wav2CLIP \cite{wav2clip}, have enabled open-vocabulary capabilities, allowing for the classification of sounds without the need for task-specific retraining. 

While generative ALMs like Qwen2-Audio \cite{qwen2} and SALMONN \cite{salmonn} have also shown promise in audio-to-text generation, contrastive learning-based models remain the cornerstone for discriminative tasks due to their cross-modal alignment. Trained on large-scale datasets \cite{laion}, CLAP leverages the HTSAT audio encoder \cite{htsat} to bridge diverse acoustic domains with their semantic captions, demonstrating state-of-the-art performance across various downstream applications \cite{enclap++, clap-art, clap-retreival, clap-sed, clap-multi-object, clap-paratask}.


To adapt these large-scale foundation models to specific downstream tasks, prompt learning has emerged as a parameter-efficient alternative to full fine-tuning~\cite{coop, cocoop}. By prepending learnable soft prompts to the input data, researchers have successfully steered models to perform better in few-shot and zero-shot scenarios. However, existing prompt learning research for ALMs~\cite{clep_dg, audiofree, palm} has been heavily text-centric. This bias stems from the inherent stability of the text embedding space, where text serves as a discrete and semantically consistent anchor, making it a reliable target for optimization. In contrast, audio signals are continuous, high-dimensional, and often exhibit significant intra-class variance and background noise, leading many to believe that optimizing the audio encoder is less efficient or more prone to instability.

We argue that this text-centric bias introduces a fundamental bottleneck. Restricting adaptation to the text modality while keeping the audio representation fixed constrains the model’s ability to achieve optimal cross-modal alignment. In this context, mutual adaptation, in which both audio and text representations are jointly refined, can be beneficial for capturing the nuanced characteristics of task-specific audio data. Text-only prompts cannot fully bridge the acoustic gap caused by domain shifts between the massive pretraining corpus and diverse downstream datasets. To the best of our knowledge, the potential of learnable prompts within the audio encoder remains unexplored in the context of few-shot audio classification.

In this paper, we propose a novel plug-in framework for \textbf{Audio-Side Prompt Learning (ASPL)} that can be seamlessly integrated with existing text-side counterparts. Our approach injects learnable prompt tokens at three strategic locations within the audio pipeline to capture task-specific acoustic features: (1) the log-mel spectrogram transformation stage, (2) the stage immediately following patch embedding, and (3) the early layers of the Swin Transformer blocks. This multi-level prompting strategy allows the model to adjust its perception of sound at both the low-level acoustic and high-level structural stages.





\section{Method}

\begin{figure*}[t] 
    \centering
    \includegraphics[width=\textwidth]{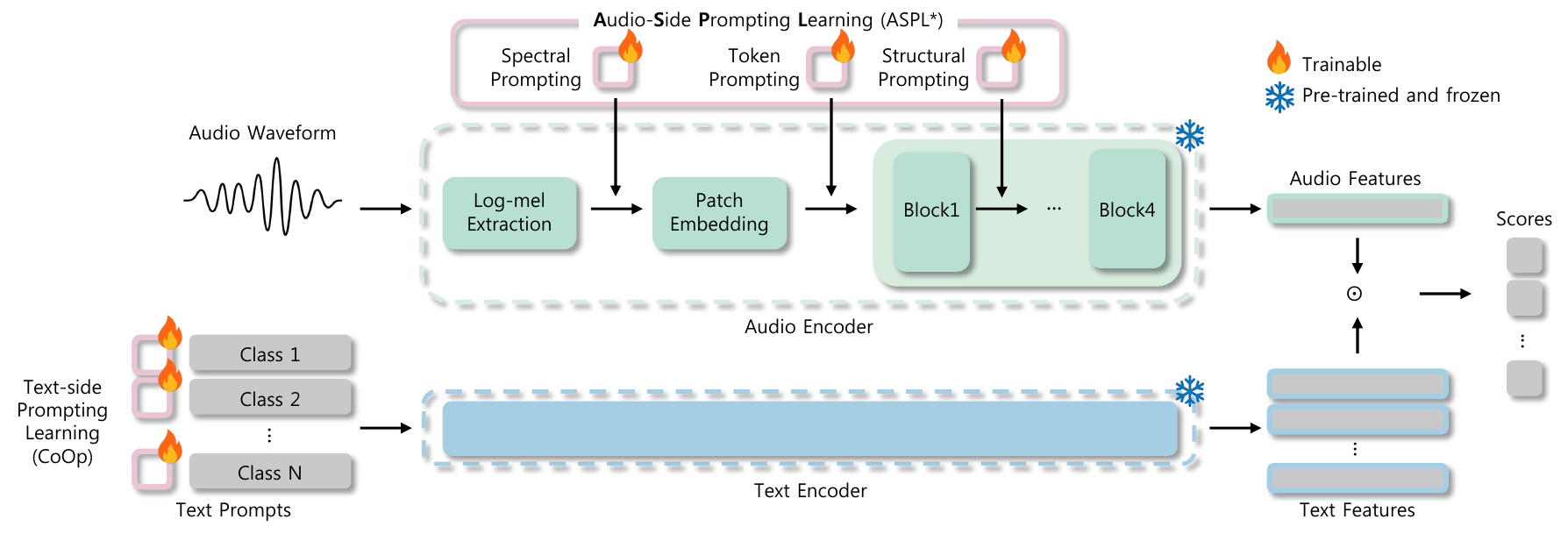} 
    \caption{\textbf{Overview of the proposed ASPL framework.} 
    Built upon the CLAP-HTSAT architecture, our method performs three-stage modulation on the audio encoder, representing the full ASPL$^*$ configuration. ASPL is designed as a \textit{plug-and-play} framework compatible with various text-side methods; while the figure illustrates integration with CoOp, it can also be applied to CoCoOp or PALM. The \textbf{fire} (\raisebox{-2pt}{\includegraphics[height=1.0em]{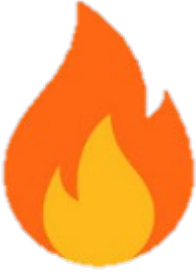}}) and \textbf{ice} (\raisebox{-2pt}{\includegraphics[height=1.0em]{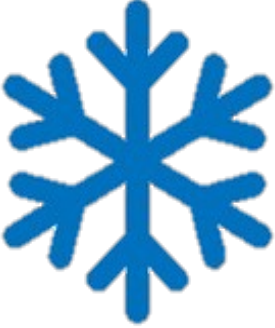}}) icons denote trainable parameters and frozen pre-trained encoders, respectively.}
    \label{fig:aspl_architecture}
\end{figure*}

\subsection{Preliminary: Audio-Text Alignment in ALMs}

Pre-trained ALMs align modality-specific representations in a shared space via cosine similarity. Given an audio waveform $x$, the audio encoder extracts a global representation $\mathbf{z}_a = E_a(x)$. For a $C$-class task, textual descriptions $\{T_i\}_{i=1}^C$ are generated using a template (e.g., \textit{``A recording of a \{class\}''}) and encoded into target embeddings $\mathbf{z}_{t,i} = E_t(T_i)$. The predicted class index $\hat{y}$ is then determined by maximizing the cosine similarity between the audio and text embeddings:
\begin{equation}
    \hat{y} = \argmax_{i \in \{1,\dots,C\}} \frac{\langle \mathbf{z}_a, \mathbf{z}_{t,i} \rangle}{\lVert \mathbf{z}_a \rVert \lVert \mathbf{z}_{t,i} \rVert},
\end{equation}
where $\langle \cdot, \cdot \rangle$ and $\lVert \cdot \rVert$ denote the inner product and $L_2$ norm, respectively.

\subsection{Text-side Prompt Learning}

Previous research has predominantly focused on optimizing the text encoder to improve cross-modal alignment. For instance, \textbf{CoOp}~\cite{coop} and \textbf{CoCoOp}~\cite{cocoop} introduce static and dynamic learnable prompt tokens, respectively. More recently, \textbf{PALM}~\cite{palm} expanded the definition of prompting beyond discrete token concatenation by demonstrating that lightweight affine transformations on text features act effectively as prompt learning. Despite their success, these methods exclusively refine text representations, leaving the potential for analogous learnable prompts within the \textit{audio encoder} unexplored.

\subsection{Audio-Side Prompt Learning (ASPL)}

Inspired by prior approaches like PALM, which successfully established affine transformations as a parameter-efficient prompting mechanism, we propose Audio-Side Prompt Learning (ASPL). As illustrated in Fig.~\ref{fig:aspl_architecture}, ASPL functions as a plug-in framework that establishes structural symmetry with text-side prompting. ASPL's core novelty lies in repurposing this affine mechanism as a stage-wise \textit{acoustic prompt}. Rather than introducing heavily parameterized adapter modules or applying arbitrary network conditioning, we specifically target the early acoustic-to-semantic transition phases: log-mel extraction, patch embedding, and the initial Swin Transformer block. This targeted placement allows the model to recalibrate low-level acoustic variations (e.g., spectral shifts) at the perceptual level. By injecting task-specific inductive biases exactly where the audio encoder builds its abstractions, we effectively prevent physical nuances from disrupting the pre-trained semantic knowledge in deeper layers.
Formally, for a 3D input feature $X$ at these targeted stages,  we apply a lightweight prompt-driven modulation:
\begin{equation}
    X' = \gamma \odot X + \beta,
\end{equation}
where $\gamma$ and $\beta$ are learnable 1D continuous prompt vectors broadcasted channel-wise across the remaining temporal or spatial dimensions. Unlike text-side methods such as PALM that require class-specific parameters, our acoustic prompt vectors are shared across all classes and instances. This architectural distinction ensures extreme parameter efficiency that remains strictly constant regardless of the dataset's label space complexity. Instead of acting merely as scaling factors, these parameters function as \textit{continuous acoustic prompts}. By optimizing these vectors in a continuous latent space, they enable the model to recalibrate spectral and structural cues for downstream tasks with minimal parameter overhead.



\subsubsection{Acoustic Prompting via Spectral Modulation}
The first ASPL stage operates directly after log-mel spectrogram $X \in \mathbb{R}^{B\times F\times T}$ (batch, frequency, time). Instead of appending discrete tokens, we introduce continuous acoustic prompts $(\gamma_{spec}, \beta_{spec}) \in \mathbb{R}^F$, which are broadcasted over $B$ and $T$ to perform frequency-wise scaling. This mechanism functions as a learnable equalizer that prompts the model to adapt to variations in energy distribution and recording conditions. By doing so, it efficiently steers the acoustic representations to adapt to spectral variations and recording conditions without requiring deep learning of acoustic bias. 

\begin{table*}[t]
\centering
\caption{\textbf{Average performance comparison on 11 datasets.} We report the average top-1 accuracy (\%) in a 16-shot setting across three independent runs (seeds 0, 1, 2). \textbf{Bold} and \underline{underlined} denote the best and second-best results, respectively. Our ASPL variants provide overall performance gains over text-side prompt learning baselines.}
\label{tab:audio_fewshot_results}
\newcolumntype{Y}{>{\centering\arraybackslash}X}
\begin{tabularx}{\textwidth}{l|YYYYYYYYYYY|Y}
\toprule         
Method 
& \scriptsize \makecell{Beijing\\Opera}
& \scriptsize \makecell{CRE\\MA-D}
& \scriptsize ESC50 
& \scriptsize \makecell{ESC50\\Actions}
& \scriptsize \makecell{GT\\Music\\Genre}
& \scriptsize \makecell{NS\\Instru\\ments}
& \scriptsize \makecell{RAV\\DESS}
& \scriptsize SESA 
& \scriptsize \makecell{TUT\\2017} 
& \scriptsize \makecell{Urban\\Sound}
& \scriptsize \makecell{Vocal\\Sound}
& \footnotesize Average \\
\midrule 
CoOp 
& 96.18 & 39.44 & 93.80 & 94.92 & 74.50 & 61.65 & 40.26 & 89.52 & 68.20 & 76.10 & 74.57 & 73.56\\
\quad + ASPL
& 96.91 & 35.12 & 94.10 & 96.25 & 76.17 & 63.68 & 42.09 & 88.89 & 74.90 & 76.30 & 74.11 & \underline{74.41}\\
\quad + ASPL$^{*}$ 
&96.75 & 37.75 & 94.35 & 95.50 & 78.00 & 65.68 & 42.77 & 91.11 & 74.38 & 77.40 & 77.28 & \textbf{75.54} \\
\midrule 
CoCoOp 
& 98.17 & 39.87 & 94.72 & 97.75 & 77.83 & 64.00 & 43.11 & 89.84 & 75.84 & 78.67 & 81.17 & 76.45 \\
\quad + ASPL
& 98.60 & 40.45 & 95.37 & 97.83 & 79.67 & 64.76 & 46.64 & 92.38 & 78.68 & 79.65 & 82.27 & \textbf{77.85} \\
\quad + ASPL$^{*}$
& 97.88 & 38.06 & 95.32 & 97.42 & 79.83 & 64.58 & 43.72 & 93.02 & 78.76 & 79.72 & 82.13 & \underline{77.31} \\
\midrule 
PALM 
& 98.74 & 36.15 & 96.33 & 98.25 & 80.33 & 66.24 & 46.16 & 90.79 & 79.80 & 80.97 & 82.64 & 77.86 \\
\quad + ASPL
& 98.59 & 38.08 & 96.52 & 97.75 & 82.17 & 66.55 & 49.76 & 92.06 & 82.51 & 81.73 & 83.03 & \underline{78.98} \\
\quad + ASPL$^{*}$
& 97.89 & 43.41 & 96.58 & 98.00 & 80.00 & 65.19 & 48.20 & 94.61 & 83.14 & 82.21 & 82.60 & \textbf{79.26} \\
\bottomrule 
\end{tabularx}
\end{table*}

\subsubsection{Token-Level Prompting at the Patch Embedding Interface}
The second ASPL stage modulates the latent tokens $X \in \mathbb{R}^{B\times L\times C}$ (batch, token length, channel) generated by the Patch Embedding layer. 
By performing channel-wise recalibration via the learnable parameters $(\gamma_{tok}, \beta_{tok}) \in \mathbb{R}^C$ broadcasted over $B$ and $L$, we adaptively emphasize task-specific acoustic patterns. This interface-level refinement seamlessly bridges the transition from local convolutional features to global attention mechanisms, potentially aligning the token space with the target domain.

\subsubsection{Structural Prompting via Early-Layer Conditioning}
The final structural modulation stage targets the output of the initial Swin Transformer block. By broadcasting parameters $(\gamma_{block1}, \beta_{block1}) \in \mathbb{R}^C$ across batch $B$ and sequence length $L$, we condition representations immediately after the first local-global mixing. This task-specific bias stabilizes early features, adapting to acoustic nuances while preventing semantic drift in deeper layers.



\subsubsection{Multi-modal Alignment and Practical Scalability}
Together, these three prompting stages systematically modulate distinct levels of acoustic abstraction—from raw spectral features to deeper structural representations. When integrated with existing text-side prompting methods (e.g., CoOp, PALM), as a plug-and-play module, ASPL forms a comprehensive dual-sided prompting framework. This jointly refines both modalities, promoting balanced cross-modal alignment rather than forcing the text encoder to adapt to a frozen, static audio space. Consequently, this prompt-driven joint optimization enhances few-shot classification, while its extreme parameter efficiency provides a scalable alternative to full fine-tuning for large foundation models.



\section{Experiments}

\subsection{Datasets}
To comprehensively evaluate the effectiveness of our proposed method, we conducted experiments on 11 diverse audio classification datasets, following the benchmark suite and evaluation protocol established in PALM. The datasets cover a wide range of tasks, including: instrument classification (Beijing-Opera \cite{beijing}, NS-Instruments \cite{ns-instruments}); sound event classification (ESC-50 \cite{esc50}, ESC-50-Action \cite{esc50}, UrbanSound8K \cite{urban}); emotion recognition (CREMA-D \cite{crema}, RAVDESS \cite{ravdess}); vocal sound classification (VocalSound \cite{vocal}); surveillance event classification (SESA \cite{sesa}); acoustic scene classification (TUT2017 \cite{tut}); and music analysis (GT-Music-Genre \cite{gt-music}). We utilized the official train-test splits where available, and for datasets without official splits, we employed a multi-fold cross-validation strategy, reporting the average accuracy.

\subsection{Implementation Details}

\noindent\textbf{Baselines and Architecture.}
We compared our proposed ASPL against state-of-the-art text-side prompt learning methods: CoOp, CoCoOp, and PALM.
Following PALM~\cite{palm}, we utilized CLAP-HTSAT as the audio encoder and the CLIP text encoder~\cite{clip} from pretrained PENGI~\cite{pengi}.
During training, both encoders were frozen, updating only the prompt parameters and ASPL modules.

\noindent\textbf{Training Settings.} All experiments followed a standard few-shot protocol, where $K \in \{1, 2, 4, 8, 16\}$ training samples were randomly selected per class. We fixed random seeds to $\{0, 1, 2\}$ for three independent runs and reported the average Top-1 accuracy. On the text-side, a single prompt context was utilized without ensembling. All models were trained for 100 epochs using cross-entropy loss and SGD (learning rate: 0.01, momentum: 0.9, weight decay: 0) with a batch size of 16 on an NVIDIA RTX A5000. Both audio and text embeddings were $L_2$-normalized, and a fixed temperature $\tau=0.01$ was applied to the cosine similarity scores prior to the softmax layer, following the established prompt optimization protocol of PALM.

\subsection{Main Results}
Table \ref{tab:audio_fewshot_results} presents the quantitative comparison across 11 datasets. We designate our approach utilizing spectral and token modulations as \textbf{ASPL}, while \textbf{ASPL$^*$} refers to the extended configuration incorporating early-layer structural conditioning. To ensure a fair comparison, our baselines focus exclusively on training-based adaptation methods.

The results demonstrate that our audio-side modulation provides a highly favorable performance-efficiency trade-off. Despite minor fluctuations per dataset, it achieves an average gain of $\sim$1.1\% with a minimal parameter overhead. Notably, the optimal modulation depth correlates with the adaptability of the base text encoder:

\begin{figure*}[th] 
    \centering
    \includegraphics[width=1\textwidth]{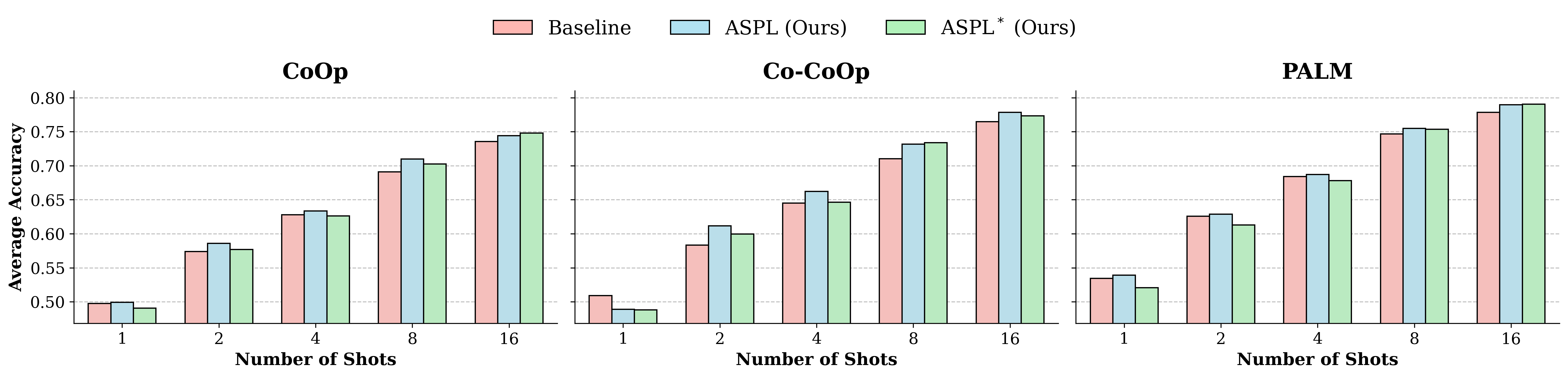}
    \caption{\textbf{Ablation study on the number of shots.} 
    Comparison of average accuracy on 11 datasets for CoOp, CoCoOp, and PALM backbones. Our proposed ASPL variants (solid bars) consistently outperform the baselines (pink bars) across most few-shot settings, demonstrating robust adaptation capabilities.}
    \label{fig:nshot}
\end{figure*}


\begin{itemize}
    \item \textbf{Synergy with Dynamic Prompts (CoCoOp):} For CoCoOp, which generates dynamic instance-conditioned text prompts, the lighter ASPL configuration achieves state-of-the-art performance. Since CoCoOp already handles instance-level variance, the focused spectral and token modulation of ASPL provides sufficient acoustic alignment without introducing redundancy.
    
    \item \textbf{Synergy with Static/Class Prompts (CoOp, PALM):} Conversely, for CoOp and PALM, which rely on static or class-specific prompts, the deeper structural modulation of ASPL$^*$ proves essential. These methods benefit significantly from the additional adaptability provided by early-layer conditioning, which compensates for the rigidity of the text representation.
\end{itemize}

In summary, our framework functions as a flexible plug-and-play module that effectively bridges the modality gap, with the deeper ASPL$^*$ serving as a powerful compensator for less flexible text backbones.

\begin{table}[t]
\centering
\vspace{-4mm}
\caption{\textbf{Complexity and Performance Comparison.} Comparison of trainable parameters, inference latency (measured on CREMA-D), and average accuracy using the PALM baseline. Integrating ASPL and ASPL$^*$ improves overall accuracy while adding minimal parameter overhead and marginal latency variations (within the noise floor across 15 trials).}
\label{tab:params}
\setlength{\tabcolsep}{9pt}
\resizebox{\columnwidth}{!}{%
\begin{tabular}{l|ccc}
\toprule
\textbf{Method} & \textbf{\# Params} & \textbf{Latency (ms)} & \textbf{Avg. Acc. (\%)} \\
\midrule

CoOp    & 8,192 & 2.60 & 73.56 \\
CoCoOp  & 107,072 & 14.76 & 76.45 \\
\midrule
PALM  & 4,100 & 2.60 & 77.86 \\
\rowcolor{lightgray}
\rowcolor{lightgray}
\quad + ASPL & 4,420 & 3.09 & 78.98 \\
\rowcolor{lightgray}
\quad + ASPL$^{*}$ & 4,804 & 2.97 & \textbf{79.26} \\

\bottomrule
\end{tabular}
}
\end{table}

\subsection{Ablation Study}

\noindent\textbf{Impact of Shot Availability.} 
Figure~\ref{fig:nshot} illustrates the performance of ASPL across varying shot counts. While we observe a slight performance degradation in the 1-shot setting—likely because a single example provides insufficient supervision to robustly optimize the continuous acoustic prompts without minor overfitting—our method consistently outperforms text-only baselines from the 2-shot setting onwards. This indicates that once a minimal threshold of data is available, our hierarchical prompting strategy effectively leverages the supervision to refine acoustic representations, yielding a steadily higher performance ceiling across 4, 8, and 16-shot setups.


\noindent\textbf{Parameter Efficiency.} A key advantage of our approach is the extreme parameter efficiency shown in Table~\ref{tab:params}. While baseline methods can introduce up to 107,072 additional parameters scaling with the number of classes, our method adds a negligible and fixed number of parameters regardless of dataset complexity: only 320 for ASPL and 704 for ASPL$^*$. This extreme efficiency is achieved by deliberately targeting the early, low-dimensional stages of the audio encoder. Specifically, the parameter count directly corresponds to the affine pairs applied to 64-mel bins (128 parameters), the 96-dimensional initial token projection (192 parameters), and for ASPL$^*$, the 192-dimensional first Transformer block (384 parameters). 
Remarkably, despite the negligible increase in parameters and inference latency, our method consistently yields an absolute accuracy gain of over 1.4\%, proving highly suitable for resource-constrained, real-time audio applications.


\noindent\textbf{Effect of Modulation Location.} Table~\ref{tab:checkmark} demonstrates that our performance gains are rooted in deliberate architectural choices rather than arbitrary prompt insertion. While individual modulations in rows 2--3 cause slight degradation—suggesting that adapting isolated components abruptly disrupts the internal representational flow of the frozen encoder—our integrated ASPL (Row 5) achieves a synergistic gain. This indicates that a coordinated, multi-stage recalibration is essential to smoothly bridge acoustic variations without causing internal covariate shifts. Importantly, simply replicating the output-space modulation used in PALM for the audio encoder (Rows 4 and 6) proves ineffective for audio signals, confirming that our success is not a byproduct of naive adaptation of text-centric methods. Furthermore, the effectiveness of early-block modulation in ASPL$^*$ (Row 8) over late-block modulation (Row 7) reinforces our core design principle that complex audio signals, unlike text, necessitate early-stage acoustic conditioning to address domain shifts while preserving pre-trained cross-modal knowledge.

\begin{table}[t]
\centering
\vspace{-4mm}
\setlength{\tabcolsep}{10pt}
\caption{\textbf{Ablation study on modulation locations} (Baseline: CoOp). Combining spectral, token, and structural modulations yields the optimal synergy for audio-text alignment.}
\label{tab:checkmark}
\resizebox{\columnwidth}{!}{%
\begin{tabular}{lccc|c}
\toprule

& \textbf{Spectral}
& \textbf{Token}
& \textbf{Structural}
& \textbf{Avg. Acc. (\%)} \\
\midrule
1 &  &  &  & 73.56 \\
2 & \checkmark &  &  & 73.35 \\
3 &  & \checkmark &  & 72.96 \\
4 &  &  & Output Space & 72.94 \\
\rowcolor{lightgray}
5 & \checkmark & \checkmark &  & \underline{74.41} \\
6 & \checkmark & \checkmark & Output Space & 73.90 \\
7 & \checkmark & \checkmark & Late Block & 74.09 \\
\rowcolor{lightgray}
8 & \checkmark & \checkmark & Early Block & \textbf{75.54} \\
\bottomrule
\end{tabular}%
}
\end{table}


\section{Conclusion}

In this paper, we introduced Audio-Side Prompt Learning (ASPL), a \textit{plug-and-play} framework designed to overcome the text-centric limitations of prompt learning in current audio-language models. By strategically injecting lightweight, coordinated modulations at the early spectral, token, and structural levels, our approach facilitates a balanced mutual adaptation without disrupting the pre-trained representational flow. Extensive evaluations across 11 diverse datasets demonstrate that ASPL improves the overall average accuracy compared to text-only baselines while maintaining remarkable parameter efficiency. These results underscore that unlocking the plasticity of the audio encoder's early stages is a vital, yet often overlooked, component for effective few-shot recognition. Moving forward, we aim to explore dynamic gating mechanisms to automatically calibrate modulation depth, further advancing the development of adaptive multi-modal foundation models.

\section{Acknowledgements}
This work was supported by IITP grants funded by the Korean government (MSIT, RS-2025-02263169,
Detection and Prediction of Emerging and Undiscovered Voice Phishing).

\section{Generative AI Use Disclosure}
The authors acknowledge the use of Google's Gemini during the drafting of this paper for the purpose of language polishing and grammatical correction. This tool was employed exclusively to improve the linguistic quality of the text and played no role in formulating the research methodology, conducting experiments, or deriving scientific insights. The authors have thoroughly validated all phrasing and hold sole responsibility for the integrity and accuracy of the submitted manuscript.

\bibliographystyle{IEEEtran}
\bibliography{shortstrings,mybib}

\end{document}